\documentclass[aps,prl,preprint,nopacs,superscriptaddress]{revtex4}

\usepackage{graphicx}
\usepackage{verbatim}
\usepackage{mathrsfs}
\usepackage{color}
\usepackage{amsmath,amsfonts,amssymb}
\usepackage{graphicx}

\def\3{2.8in}    
\def\2{2.5in}
\def\4{3.0in}

\def\nn{\nonumber}

\def \beq {\begin{equation}}
\def \eeq {\end{equation}}
\begin{document}

\title{Direct optical detection of Weyl fermion chirality in a topological semimetal}
\author{Qiong Ma$^*$}\affiliation {Department of Physics, Massachusetts Institute of Technology, Cambridge, Massachusetts 02139, USA}

\author{Su-Yang Xu\footnote{These authors contributed equally to this work.}}\affiliation{Department of Physics, Massachusetts Institute of Technology, Cambridge, Massachusetts 02139, USA}

\author{Ching-Kit Chan}\affiliation {Department of Physics, Massachusetts Institute of Technology, Cambridge, Massachusetts 02139, USA}\affiliation{Department of Physics and Astronomy, University of California Los Angeles, Los Angeles, California 90095, USA}

\author{Cheng-Long Zhang}
\affiliation{International Center for Quantum Materials, School of Physics, Peking University, China}\affiliation{Collaborative Innovation Center of Quantum Matter, Beijing,100871, China}

\author{Guoqing Chang}
\affiliation{Centre for Advanced 2D Materials and Graphene Research Centre National University of Singapore, 6 Science Drive 2, Singapore 117546}
\affiliation{Department of Physics, National University of Singapore, 2 Science Drive 3, Singapore 117542}

\author{Yuxuan Lin}
\affiliation{Department of Electrical Engineering and Computer Science, Massachusetts Institute of Technology, Cambridge, Massachusetts 02139, USA}
\author{Weiwei Xie}
\affiliation{Department of Chemistry, Louisiana State University, Baton Rouge, Louisiana 70803-1804, USA}
\author{Tom\'as Palacios}
\affiliation{Department of Electrical Engineering and Computer Science, Massachusetts Institute of Technology, Cambridge, Massachusetts 02139, USA}

\author{Hsin Lin}
\affiliation{Centre for Advanced 2D Materials and Graphene Research Centre National University of Singapore, 6 Science Drive 2, Singapore 117546}
\affiliation{Department of Physics, National University of Singapore, 2 Science Drive 3, Singapore 117542}

\author{Shuang Jia}
\affiliation{International Center for Quantum Materials, School of Physics, Peking University, China}\affiliation{Collaborative Innovation Center of Quantum Matter, Beijing,100871, China}

\author{Patrick A. Lee}\affiliation {Department of Physics, Massachusetts Institute of Technology, Cambridge, Massachusetts 02139, USA}
\author{Pablo Jarillo-Herrero$^{\dag}$}\affiliation {Department of Physics, Massachusetts Institute of Technology, Cambridge, Massachusetts 02139, USA}

\author{Nuh Gedik\footnote{Corresponding authors (emails): pjarillo@mit.edu, and gedik@mit.edu }}\affiliation {Department of Physics, Massachusetts Institute of Technology, Cambridge, Massachusetts 02139, USA}

\date{\today}

\begin{abstract}

A Weyl semimetal (WSM) is a novel topological phase of matter \cite{Weyl, McEuen, Volovik2003, Murakami2007, Huang2015, Weng2015, Wan2011, Burkov2011, Hasan_Commentary, Hasan_TaAs, MIT_Weyl, TaAs_Ding, Nonlocal}, in which Weyl fermions (WFs) arise as pseudo-magnetic monopoles in its momentum space. The chirality of the WFs, given by the sign of the monopole charge, is central to the Weyl physics, since it directly serves as the sign of the topological number \cite {Wan2011, Hasan_Commentary} and gives rise to exotic properties such as Fermi arcs \cite {Wan2011, Hasan_TaAs, TaAs_Ding} and the chiral anomaly \cite{Hasan_Commentary, Nonlocal, Chiral_anomaly_Jia, Chiral_anomaly_ChenGF, Ong_Chiral}. Despite being the defining property of a WSM, the chirality of the WFs has never been experimentally measured. Here, we directly detect the chirality of the WFs by measuring the photocurrent in response to circularly polarized mid-infrared light. The resulting photocurrent is determined by both the chirality of WFs and that of the photons. Our results pave the way for realizing a wide range of theoretical proposals \cite{Hasan_Commentary, Nonlocal, Inti, PC_Weyl_Chris,PC_Weyl_Nagaosa,PC_Weyl_Tanaka,PC_Weyl_Moore,Weyl_Floquet,Burkov,Hosur,Pallab, Pesin} \color{black} for studying and controlling the WFs and their associated quantum anomalies by optical and electrical means. More broadly, the two chiralities, analogous to the two valleys in 2D materials \cite{Heinz_review, Mak_review}, lead to a new degree of freedom in a 3D crystal with potential novel pathways to store and carry information.

\vspace{0.6cm}

\end{abstract}
\maketitle

\begin{figure*}[t]
\includegraphics[width=16cm]{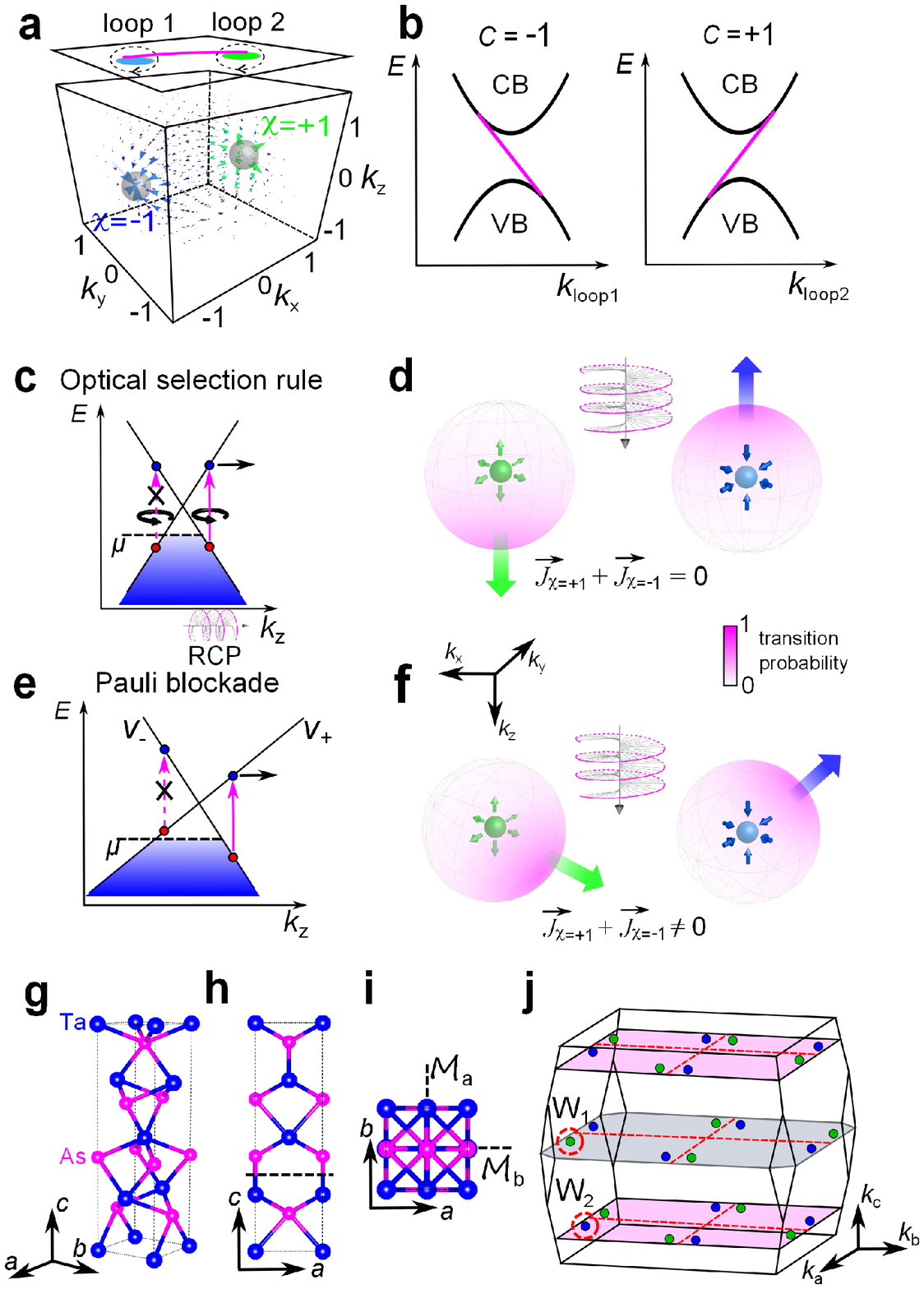}
\caption{{\bf Chirality-dependent optical transition of Weyl fermions in TaAs.}}
\label{Fig1}
\end{figure*}

In 1929, H. Weyl discovered that all elementary fermions that have zero mass must attain a definitive chirality determined by whether the directions of spin and motion are parallel or anti-parallel \cite{Weyl}. Such a chiral massless fermion is called the Weyl fermion (WF). Although none of the fundamental particles in high-energy physics was identified as WFs, condensed matter researchers have found an analog of this elusive particle in a new class of topological materials, the Weyl semimetal (WSM) \cite{McEuen, Volovik2003, Murakami2007, Huang2015, Weng2015, Wan2011, Burkov2011, Hasan_Commentary, Hasan_TaAs, MIT_Weyl, TaAs_Ding, Nonlocal}. Similar to the case in high-energy physics, the WFs in a WSM also have a definitive chirality. A right-handed Weyl node ($\chi=+1$) is a monopole (a source) of Berry curvature whereas a left-handed Weyl node ($\chi=-1$) is an anti-monopole (a drain) of Berry curvature. Any Fermi surface enclosing a right(left)-handed Weyl node ($\chi=\pm1$) has a unit Berry flux coming out (in) and hence carries a Chern number $C=\pm1$ (Fig.~\ref{Fig1}\textbf{a}). As a result, the chirality of the WF serves as its topological number. 

The distinct chirality directly leads to exotic, topologically protected phenomena in a WSM. First, the separation between WFs of opposite chirality in $k$ space protects them from being gapped out \cite{Wan2011}. Second, the opposite Chern numbers of the bulk Fermi surfaces guarantee the existence of topological Fermi arc surface states \cite{Wan2011} that connect between pairs Weyl nodes of opposite chirality (Figs.~\ref{Fig1}\textbf{a,b}). Third, applying parallel electric and magnetic fields can break the apparent conservation of chirality, making a Weyl metal, unlike ordinary non-magnetic metals, more conductive with an increasing magnetic field \cite{Chiral_anomaly_Jia, Chiral_anomaly_ChenGF, Ong_Chiral}. Besides its fundamental importance in the topological physics of WSM, the chirality also gives rise to a new degree of freedom in 3D materials, analogous to the valley degree of freedom in the 2D transition metal dichalcogenides (TMDs) that have gathered great attention recently \cite{Heinz_review, Mak_review}. The potential to control the chirality \cite{Hasan_Commentary, Nonlocal, Inti, PC_Weyl_Chris,PC_Weyl_Nagaosa,PC_Weyl_Tanaka,PC_Weyl_Moore,Weyl_Floquet,Burkov,Hosur,Pallab}, combined with the high electron mobility found in the WSMs \cite{Chiral_anomaly_Jia}, may offer new schemes to encode and process information. 

Therefore, it is of crucial importance to detect the chirality of the WFs. This requires identifying physical observables that are sensitive to the WF chirality. The band structures, quasi-particle interferences, magneto-resistances measured by angle-resolved photoemission spectroscopy (ARPES) \cite{Hasan_TaAs, TaAs_Ding}, scanning tunneling microscope \cite{STM_Hao, STM_Yazdani, STM_Haim}, and transport experiments \cite{Chiral_anomaly_Jia, Chiral_anomaly_ChenGF, Ong_Chiral}, respectively, are not sensitive to the chirality of WFs. One proposal to detect the chirality is to use pump-probe ARPES to measure the transient spectral weight upon shining circularly polarized pump light  \cite{CD_ARPES}. However, this requires an ARPES with a mid-infrared pump and a soft X-ray probe, which is technically very challenging. On the other hand, optical experiments on WSMs have remained very limited \cite{TaAs_PRB, SHG}, although they are promising approaches to achieve these goals \cite{Hasan_Commentary}. In this paper, we detect the chirality of the WFs in the WSM TaAs by measuring its mid-infrared photocurrent response. Circularly polarized light induced photocurrents, also called the circular photogalvanic effect (CPGE), have been previously measured in other systems \cite{James, Yi, Ganichev} but have not been experimentally studied in WSMs.

We first discuss the theoretical picture of the CPGE for optical transitions from the lower part of the Weyl cone to the upper part \cite{PC_Weyl_Chris}. There are two independent factors important for the CPGE here. The first is the chirality selection rule (Figs.~\ref{Fig1}\textbf{c,d}). For a right circularly polarized (RCP) light propagating along $+\hat{z}$ and a $\chi=+1$ WF, the optical transition is allowed on the $+k_z$ side but forbidden on the $-k_z$ side \cite{PC_Weyl_Chris}. The second is the Pauli blockade, which is only present when chemical potential is away from the Weyl node. In the presence of a finite tilt (Fig.~\ref{Fig1}\textbf{e}), the Pauli blockade becomes asymmetric about the nodal point. If we only consider a single Weyl cone, in general we expect a nonzero current (Figs.~\ref{Fig1}\textbf{c,d}). However, having a nonzero total photocurrent depends on whether contributions from different WFs cancel each other. In an inversion-breaking WSM with mirror symmetries, Ref. \cite{PC_Weyl_Chris} shows that the total photocurrent becomes nonvanishing when both factors are present (see also supplemental information (SI) SI.III.1). 

TaAs has been experimentally established as an inversion-breaking WSM with mirror symmetries \cite{Weng2015, Huang2015, Hasan_TaAs, TaAs_Ding}. The tilt of the WFs is significant (e.g., $\arrowvert\frac{v_+}{v_-}\arrowvert\geq2$, see Fig.~\ref{Fig1}\textbf{e} and SI.II.2) and the chemical potential is $\sim18$ meV away from the W$_1$ Weyl nodes (SI.II.1). These factors make TaAs a promising system to observe the WF-induced CPGE as discussed above. We describe the following properties of TaAs relevant for our study. While TaAs has twenty-four WFs, only two are independent, which are highlighted in Fig.~\ref{Fig1}\textbf{j} and named W$_1$ and W$_2$. The other twenty-two Weyl nodes can be related by the system's symmetries (see SI.I), including time-reversal ($\mathcal{T}$), four-fold rotation around $\hat{c}$ ($C_{4c}$) and two mirror reflections about the $(b,c)$ and $(a,c)$ planes ($\mathcal{M}_a$ and $\mathcal{M}_b$). Moreover, since the crystal lacks mirror symmetry $\mathcal{M}_c$, $+\hat{c}$ can be unambiguously defined as the direction going from the Ta atom to the As atom across the dotted line in Fig.~\ref{Fig1}\textbf{h}. With this well-defined lattice (Figs.~\ref{Fig1}\textbf{g-i}) as an input, first-principles calculations have predicted the energy-dispersion and the chirality of each WF (Fig.~\ref{Fig1}\textbf{j}). While the energy-dispersion has been observed by ARPES, the chirality configuration (Fig.~\ref{Fig1}\textbf{j}) has not been measured.

\color{black} 
In order to detect the CPGE, we utilize a mid-infrared scanning photocurrent microscope (Fig.~\ref{Fig2}\textbf{a}) equipped with a $\mathrm{CO}_2$ laser source  ($\lambda_{\textrm{CO}_2} = 10.6$ $\mu\mathrm{m}$ and $\hbar\omega\simeq 120$ meV). We note that this photon energy fits our purpose because we specifically want to excite electrons from the lower part of Weyl cone to the upper part. A much lower photon energy (in the terahertz range) will likely fall into the intra-band regime unless the chemical potential is tuned very close to the Weyl nodes, whereas a much higher photon energy would excite electrons to bands at much higher energy, making the process of marginal relevance to the Weyl physics. Throughout the paper, we assign the propagation of the light as $+\hat{x}$. Our TaAs sample (Fig.~\ref{Fig2}\textbf{b}) is purposely filed down so that the out-of-plane direction is $\hat{a}$. We have performed single crystal x-ray diffraction (XRD, see SI.II.4), which allows us to determine the $+\hat{c}$ direction (Fig.~\ref{Fig2}\textbf{b}). Throughout Fig.~\ref{Fig2}, the lab and sample coordinates are identical (e.g., $+\hat{a}=+\hat{x}$). The black data-points in Fig.~\ref{Fig2}\textbf{c} show the current along $b$, when the laser-spot is near the sample's center (the black dot in Fig.~\ref{Fig2}\textbf{b}). We observe that the current reaches maximum value for RCP light, minimum for LCP light, and zero for linearly polarized light. The whole data-curve fits nicely to a cosine function. In sharp contrast, we see no observable current along the $c$ direction (the black data-points in Fig.~\ref{Fig2}\textbf{d}).

We move the light spot horizontally to the blue and pink dots in Fig.~\ref{Fig2}\textbf{b}. The corresponding photocurrents (the blue and pink data-points in Fig.~\ref{Fig2}\textbf{c}) show the same polarization-dependence but with an additional, polarization-independent shift. We also see the same polarization-independent shift for the currents along $c$ (Fig.~\ref{Fig2}\textbf{d}). These data reveal two distinct mechanisms for photocurrent generation. To understand the polarization-independent component, in Fig.~\ref{Fig2}\textbf{e}, we show the photocurrent along $b$ with a fixed polarization (RCP), while the laser spot is varied in $(y,z)$ space. We see the current flowing to the opposite directions depending on whether the light spot is closer to the left or right contact. This spatial dependence shows that the polarization-independent component arises from the photo-thermal effect \cite{HC_Gabor, James, Yi}. Figure~\ref{Fig2}\textbf{f} shows the photocurrent as a function of polarization and the $z$ position of the laser spot, where both the polarization-dependent and polarization-independent components can be seen. In order to separate these two components, we Fourier transform from the polarization angle space to the frequency space. As shown in Fig.~\ref{Fig2}\textbf{g}, aside from the low-frequency intensities that correspond to the polarization-independent photo-thermal current, we observe a sharp peak exclusively at the frequency of $1/\pi$. This peak disappears at the $z$ values outside the sample, confirming that the observed current is the sample's intrinsic property. We show the frequency-filtered photocurrent only at the frequency of $1/\pi$. The RCP (Fig.~\ref{Fig2}\textbf{h}) light induced photocurrent is along the $-y$ direction irrespective of the location of the laser. This spatial configuration is different from that of the photo-thermal effect (Fig~\ref{Fig2}\textbf{e}). In SI.IV, we further show the temperature and laser power dependences of the photocurrent. These systematic polarization, position, temperature and power dependent measurements further confirm the CPGE and isolate it from the photo-thermal effect.

We now present two important characteristics of the observed photocurrent. The first is the cancellation of photocurrent along certain directions. For RCP light along $\hat{a}$ (Fig.~\ref{Fig2}), we observe zero current along $\hat{c}$. On another TaAs sample (SI.IV), we shine light along $\hat{c}$. We observe zero current along both $\hat{a}$ and $\hat{b}$ directions. The second is the sign-reversal of the photocurrent upon rotating the sample by 180$^\circ$ while fixing the properties of the light (Fig.~\ref{Fig3}). The lab coordinate remains while the sample coordinate changes upon rotations. $I>0$ is consistently defined as $-\hat{y}$ in the lab coordinate. For a given polarization, the direction of the photocurrent reverses if one rotates the sample by $180^{\circ}$ around $\hat{a}$ or $\hat{b}$ (Figs.~\ref{Fig3}\textbf{b,c}), while it remains the same upon a rotation around $\hat{c}$ (Fig.~\ref{Fig3}\textbf{d}). To explain these observations, we consider the second order photocurrent response tensor $\eta_{\alpha \beta \gamma}$ since our CPGE shows a linear dependence to the laser power (SI.IV). $\eta_{\alpha \beta \gamma}$ is defined through:
\begin{eqnarray}
J_\alpha &=& \eta_{\alpha \beta \gamma}E_\beta(\omega) E^*_\gamma(\omega),
\end{eqnarray}
where $J$ is the total photocurrent and $E$ is the electric field. The CPGE corresponds to the imaginary part of $\eta_{\alpha \beta \gamma}$ \cite{Ganichev}. Because the $\eta_{\alpha \beta \gamma}$ tensor is an intrinsic property, it has to obey the symmetries of the system. Therefore, symmetry dictates many important properties of the CPGE, independent of the details of the band structure, the wavelength of the light, or the underlying microscopic mechanism for the optical transition. In TaAs, the presence of $\mathcal M_a$ and $\mathcal M_b$ forces $\eta_{\alpha \beta \gamma}$ to vanish when it contains an odd number of momentum index $a$ or $b$. In SI.III.2, we show that both observations can be explained by symmetry analysis of the tensor $\eta_{\alpha \beta \gamma}$, which further confirms the intrinsic nature of the observed CPGE.

While the cancellation and sign-reversal are solely dictated by symmetry, the absolute direction of the current, i.e., the sign of $\eta_{bbc}$ and $\eta_{bcb}$, depends on the microscopic mechanism. Specifically, when we shine light along $\hat{a}$, symmetry tells us the current will be nonzero along $\hat{b}$ but it cannot decide whether the current flows to $+\hat{b}$ or $-\hat{b}$ (Fig.~\ref{Fig2}). Unlike a high photon energy which would likely involve many bands irrelevant to the WFs, the photon energy $120$ meV creates excitations from the lower part of the Weyl cone to the upper part, which directly relies on the chirality of the WFs (Fig.~\ref{Fig1}). Moreover, the microscopic theory describing this optical transition is available \cite{PC_Weyl_Chris}, which allows us to theoretically calculate the photocurrent. Specifically, the photocurrent from a single WF is given by

\begin{align}
J_{\textrm{single WF}} &=  C\bar{J} \nn\\
\bar{J} & = \int{d^3q}[v_+(\vec{q})-v_-(\vec{q})]\arrowvert{V_{+-}(\vec{A}, \chi)}\arrowvert^2\delta(E_+(\vec{q})-E_-(\vec{q}) -\hbar \omega) [n^0_{-}(\mu, \vec{q})-n^0_{+}(\mu, \vec{q})],
\end{align}

where $C$ is a constant determining the magnitude of the current that is the same for all 24 WFs, $\bar{J}$ is a dimensionless vector that gives the direction of the current, $\vec{q}$ is the momentum vector from the Weyl node, $E_{\pm}$, $v_{\pm}$, and $n^0_{\pm}(\mu, \vec{q})$ are the energies, group velocities, and equilibrium distribution functions for the upper ($+$) and lower ($-$) parts of the Weyl cone (the Pauli blockade sets in through the chemical potential $\mu$ in $n^0_{\pm}$), $\omega$ and $\vec{A}$ are the frequency and vector potential of the light, and $V_{+-}$ is the optical transition matrix element, which depends on the chirality selection rule through $\chi$ and $\vec{A}$. Interestingly, the chemical potential of TaAs is known from previous transport experiments \cite{Chiral_anomaly_Jia, Chiral_anomaly_ChenGF} to be very close to the W$_2$ Weyl node. Our quantum oscillation measurements (SI.II.1) confirmed this case ($\mu$ is 17.8 meV above W$_1$ and 4.8 meV above W$_2$). Since the Pauli blockade vanishes for chemical potential at the Weyl node, the photocurrent contribution from W$_2$ WFs is negligibly small. Therefore, our photocurrent solely probes the chirality of a single W$_1$ WF (the other 7 W$_1$ WFs are automatically related by symmetries). Assuming RCP light propagating along $+\hat{a}$, this theory predicts a photocurrent ($\hat{J}_{\textrm{THY}}$) along $\hat{b}$ given by (SI.II.3)

\begin{equation}
\hat{J}_{\textrm{THY}}=\chi_{\textrm{W}_1}\hat{k}_{\textrm{light}}\times\hat{c},
\label{THY}
\end{equation} 

where $\chi_{\textrm{W}_1}$ is the chirality of the W$_1$ WF highlighted in Fig.~\ref{Fig4}\textbf{a}. In other words, theory predicts that, if $\chi_{\textrm{W}_1}=+1$, then $\hat{J}_{\textrm{THY}}$ is along $-\hat{b}$ and that if $\chi_{\textrm{W}_1}=-1$, then $\hat{J}_{\textrm{THY}}$ is along $+\hat{b}$. In our data (Fig.~\ref{Fig4}\textbf{b}), because we shine RCP light along $+\hat{a}$ and measure a current flowing to $-\hat{b}$ and because we have measured $+c$ from single crystal XRD, we observe the following relation from our data:

\begin{equation}
\hat{J}_{\textrm{EXP}}=\hat{k}_{\textrm{light}}\times\hat{c},
\label{EXP}
\end{equation} 

By comparing Eq.~\ref{THY} with Eq.~\ref{EXP}, we determine $\chi_{\textrm{W}_1}=+1$. We find that the chirality of the W$_1$ WF determined by the photocurrent agrees with that of predicted by first-principles (Fig.~\ref{Fig1}\textbf{j}) agree. The agreement further confirms our detection of the WF chirality in TaAs.

\color{black} 
Finally, we discuss how our results can open up new experimental possibilities for studying and controlling the WFs and their associated quantum anomalies. Analogous to two valleys in 2D (Fig.~\ref{Fig4}\textbf{c}), the key is to identify ways to interact with the two chiralities in a WSM distinctly. In the present study, this is achieved because the circularly polarized light excites opposite sides of the WFs of opposite chirality (Fig.~\ref{Fig4}\textbf{d}). Another approach to differentiate the two chiralities is to create a population imbalance (Fig.~\ref{Fig4}\textbf{e}). Interestingly, doing so in a WSM fundamentally requires breaking the apparent conservation of chirality (the chiral anomaly). This can be achieved electrically by applying parallel electric and magnetic fields (Fig.~\ref{Fig4}\textbf{e} \cite{Nonlocal, Chiral_anomaly_Jia, Chiral_anomaly_ChenGF, Ong_Chiral}, or optically through the chiral magnetic effect by shining a light on a special kind of WSM, where Weyl nodes of opposite chirality have different energies \cite{PC_Weyl_Tanaka, Burkov, Hosur, Pallab, Pesin}. Apart from the present study and the proposed anomaly related physics \cite{Nonlocal, PC_Weyl_Tanaka, Burkov, Hosur, Pallab, Pesin}, the nontrivial Berry curvatures in WSMs can also lead to various novel optical phenomena such as photocurrents \cite{PC_Weyl_Nagaosa, PC_Weyl_Moore, Inti}, Hall voltages \cite{Weyl_Floquet}, Kerr rotations \cite{PC_Weyl_Moore}, and second harmonic generations \cite{PC_Weyl_Moore, Inti}. While these novel phenomena \cite{PC_Weyl_Tanaka, Burkov, Hosur, Pallab, Pesin, Nonlocal, PC_Weyl_Nagaosa, PC_Weyl_Moore, Inti, Weyl_Floquet} may arise from different aspects of Weyl physics, they are ultimately rooted in the existence of two distinct chiralities ($\chi=\pm1$) WFs as demonstrated here.

\vspace{2cm}

\textbf{Methods}

\textbf{Mid-infrared photocurrent microscopy setup:} In our experiment, the sample is contacted with metal wires and placed in an optical scanning microscope setup that combines electronic transport measurements with light illumination \cite{HC_Gabor, PTE_Herring}. The laser source is a temperature-stablized $\mathrm{CO}_2$ laser with a wavelength $\lambda = 10.6$ $\mu\mathrm{m}$ ($\hbar\omega\simeq 120$ meV). A focused beam spot (diameter $d$ $\approx$ 50 $\mu$m) is scanned (using a two axis piezo-controlled scanning mirror) over the entire sample and the current is recorded at the same time to form a colormap of photocurrent as a function of spatial positions. Reflected light from the sample is collected to form a simultaneous reflection image of the sample. The absolute location of the photo-induced signal is therefore found by comparing the photocurrent map to the reflection image. The light is first polarized by a polarizer and the chirality of light is further modulated by a rotatable quarter-wave plate characterized by an angle $\theta$ (Fig.~\ref{Fig2}\textbf{a}). 

\textbf{Single crystal growth:} Single crystals of TaAs were prepared by the standard chemical vapor transfer (CVT) method \cite{Murray}. The polycrystalline samples were prepared by heating up the stoichiometric mixtures of high quality Ta (99.98\%) and As (99.999\%) powders in an evacuated quartz ampoule. Then the powder of TaAs (300 mg) and the transport agent were sealed in a long evacuated quartz ampoule (30 cm). The end of the sealed ampoule was placed horizontally at the center of a single-zone furnace. The central zone of the furnace was slowly heated up to 1273 K and kept at the temperature for 5 days, while the cold end was less than 973 K. Magneto-transport measurements were performed using a Quantum Design Physical Property Measurement System.

\textbf{First-principles calculations:} First-principles calculations were performed by the OPENMX code within the framework of the generalized gradient approximation of density functional theory \cite{Perdew}. Experimental lattice parameters were used, and the details for the computations can be found in Ref. \cite{Murray}. A real-space tight-binding Hamiltonian was obtained by constructing symmetry-respecting Wannier functions for the As $p$ and Ta $d$ orbitals without performing the procedure for maximizing localization.

\vspace{1cm}
\textbf{Acknowledgement:} N.G and S.X acknowledge support from U.S. Department of Energy, BES DMSE, award number DE-FG02-08ER46521 (initial planning), the Gordon and Betty Moore FoundationÕs EPiQS Initiative through Grant GBMF4540 (data analysis),  and in part from the MRSEC Program of the National Science Foundation under award number DMR - 1419807 (data taking and manuscript writing). Work in the PJH group work was partly supported by the Center for Excitonics, an Energy Frontier Research Center funded by the US Department of Energy (DOE), Office of Science, Office of Basic Energy Sciences under Award Number DESC0001088 (fabrication and measurement) and partly through AFOSR grant FA9550-16-1-0382 (data analysis), as well as the Gordon and Betty Moore Foundation's EPiQS Initiative through Grant GBMF4541 to PJH. This work made use of the Materials Research Science and Engineering Center Shared Experimental Facilities supported by the National Science Foundation (NSF) (Grant No. DMR-0819762). PAL acknowledge the support by DOE under grant DE-FG02-03-ER46076. TP and YL acknowledge the partial funding support from the ONR PECASE project and the MIT/Army Institute for Soldier Nanotechnologies. First-principles band structure calculations done in HL were supported by the National Research Foundation (NRF), Prime Minister's Office, Singapore, under its NRF fellowship (NRF Award No. NRF-NRFF2013-03). Single-crystal growth was supported by National Basic Research Program of China (grant Nos. 2013CB921901 and 2014CB239302).
 
\vspace{1cm}
\textbf{Author contributions:} All authors contributed to this work.

\vspace{1cm}
\textbf{Competing financial interests:} The authors declare no competing financial interests.

\addtocounter{figure}{-1}
\begin{figure*}[t!]
\caption{\textbf{a,} The blue and green arrows depict the Berry curvatures in momentum space. The grey spheres represent the Fermi surfaces that enclose the Weyl nodes. \textbf{b,} Surface band structures along the close $k$ loops in the surface Brillouin zone (BZ) defined by the dashed circles in panel (\textbf{a}). \textbf{c,} Chirality selection rule: A right-handed circularly polarized (RCP) light along $+\hat{z}$ excites the $+k_z$ side of the $\chi$ = +1 Weyl node but the $-k_z$ side of the $\chi$ = -1 Weyl node. Chirality selection rule is independent of the tilt of the WFs. \textbf{d,} In the presence of a finite tilt and a finite chemical potential away from the Weyl node, the Pauli blockade becomes asymmetric about the nodal point. \textbf{e,f,} Optical excitation scheme for a pair of un-tilted or tilted Weyl cones. The colormap on the sphere schematically shows the magnitude of the matrix element (i.e., the transition probability considering both the chirality selection rule and the Paul blockade) at different solid angles. The currents from a pair of untilted WFs of opposite chirality cancel while those from a pair of tilted WFs do not necessarily cancel. \textbf{g-i,} Crystal structure of TaAs shown from different perspectives. \textbf{j,} Distribution of the 24 Weyl nodes in the BZ of TaAs. We denote the 8 Weyl nodes in the $k_z = 0$ plane (grey) as W$_1$ and the 16 Weyl nodes in the upper and lower $k_z$ planes (pink) as W$_2$. The green and blue colors denote positive and negative chiralities, respectively.}
\end{figure*}

\begin{figure*}[t]
\includegraphics[width=17cm]{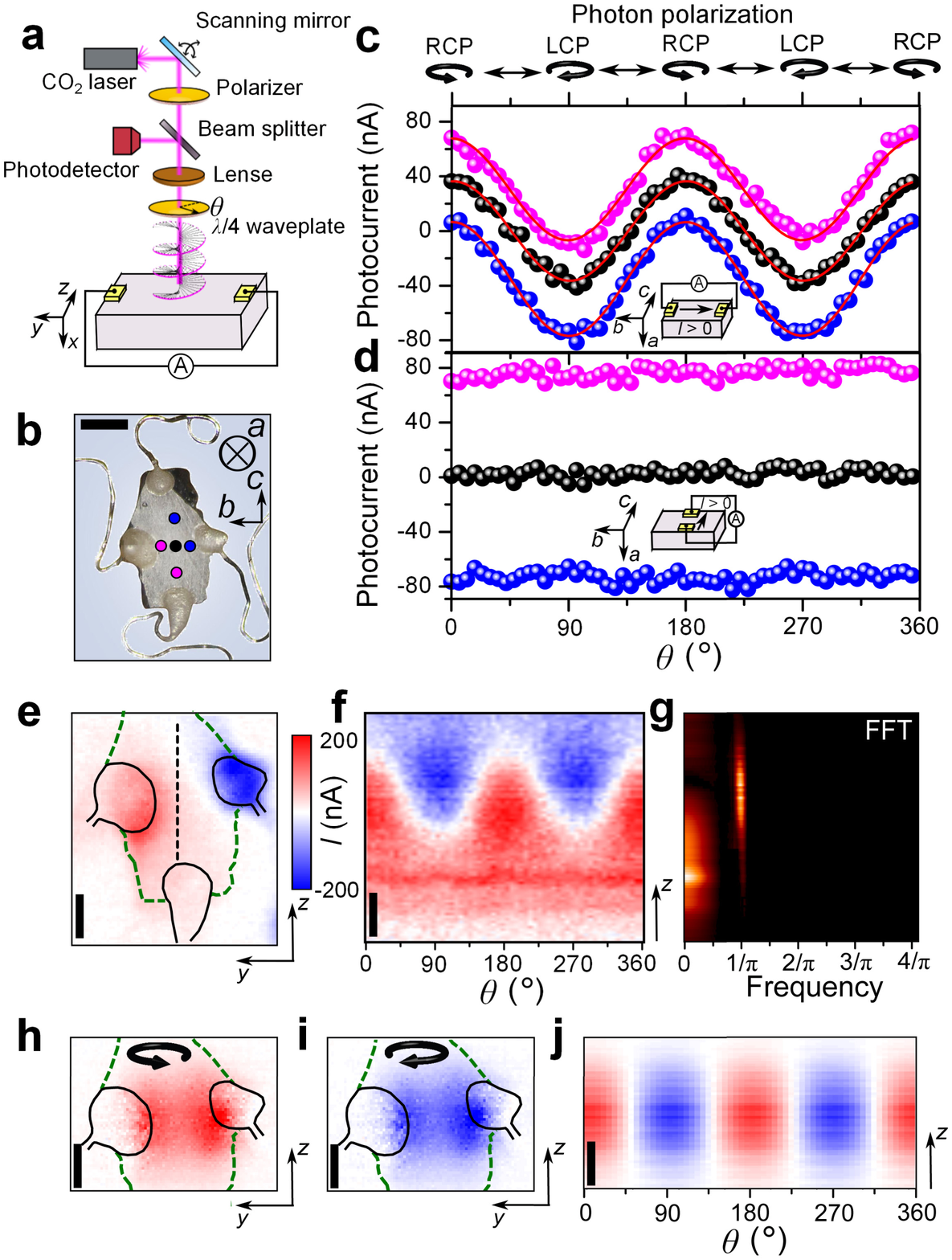}
\caption{
{\bf Observation of chirality-dependent photocurrent in TaAs.}}
\label{Fig2}
\end{figure*}
\addtocounter{figure}{-1}
\begin{figure*}[t!]
\caption{\textbf{a,} Schematic illustration of the mid-IR photocurrent microscope setup. We used a laser power about $10$ mW throughout the maintext. \textbf{b,} A photograph of the measured TaAs sample. The crystal axes $\hat{a}$, $\hat{b}$, $\hat{c}$ are denoted. Scale bar: $300$ $\mu{m}$. \textbf{c,d,} Polarization-dependent photocurrents measured along the $\hat{y}$ (panel (\textbf{c})) or $\hat{z}$ (panel (\textbf{d})) direction with the laser applied at the horizontally (panel (\textbf{c})) or vertically (panel (\textbf{d})) aligned pink, black and blue dots in panel (\textbf{b}). \textbf{e,} The color map shows the photocurrent measured along $b$ with a fixed polarization (RCP) while the laser spot is varied in $(y,z)$ space. Contacts and sample edges are traced by black solid and green dashed lines, respectively. \textbf{f,} Photocurrent as a function of polarization and the $z$ position of the laser spot. The $y$ position of the laser spot is fixed at the dashed line in panel (\textbf{e}). \textbf{g,} The Fourier transform of panel (\textbf{e}) from polarization angle ($\theta$) space to frequency ($f$) space. \textbf{h-j,} Frequency-filtered photocurrents only at the frequencies of $1/\pi$.}
\end{figure*}

\clearpage
\begin{figure*}[t]
\includegraphics[width=17cm]{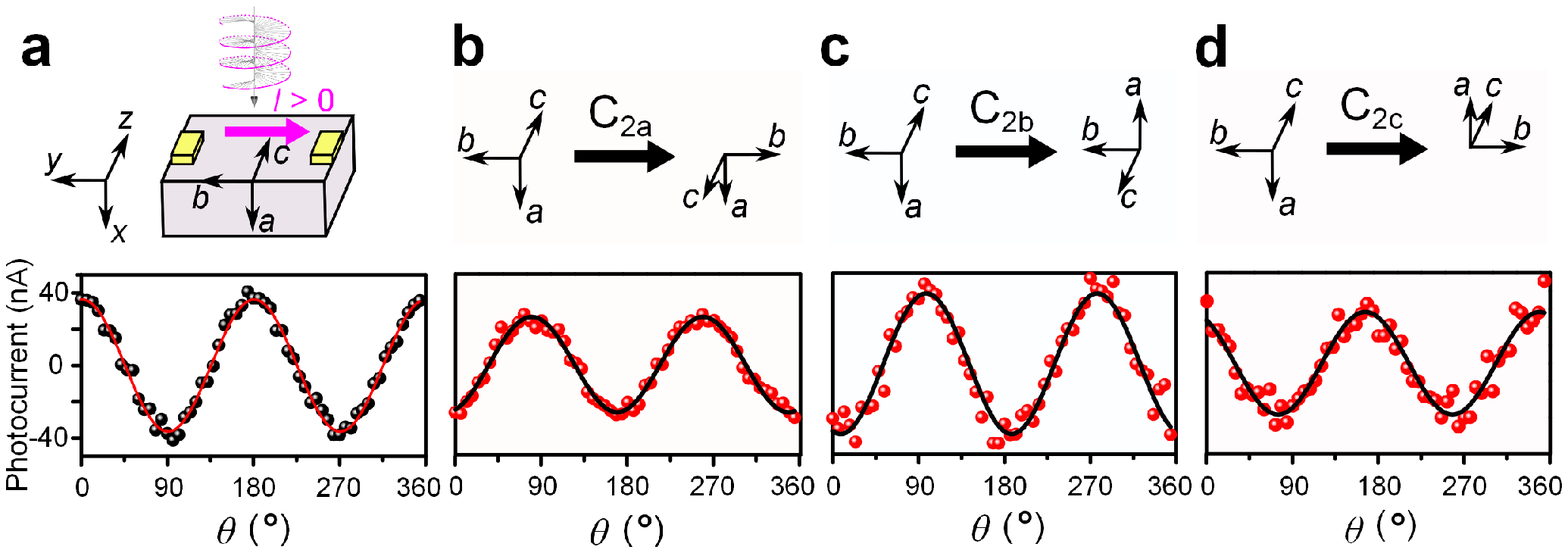}
\caption{
{\bf Control of photocurrent by varying the Weyl fermion chirality configuration with respect to the light.} \textbf{a,} Repetition of Fig. 2\textbf{c}. Both the laboratory coordinate ($\{\hat{x}, \hat{y}, \hat{z}\}$) and the sample coordinate ($\{\hat{a}, \hat{b}, \hat{c}\}$) are noted. $I>0$ is defined as along the -$\hat{y}$ direction in the laboratory framework. \textbf{b-d,} Photocurrent measurements after a $180^{\circ}$ rotation of the sample around $\hat{a}$, $\hat{b}$ or $\hat{c}$ ($C_{2a}$, $C_{2b}$, or $C_{2c}$). }
\label{Fig3}
\end{figure*}

\begin{figure*}[t]
\includegraphics[width=17cm]{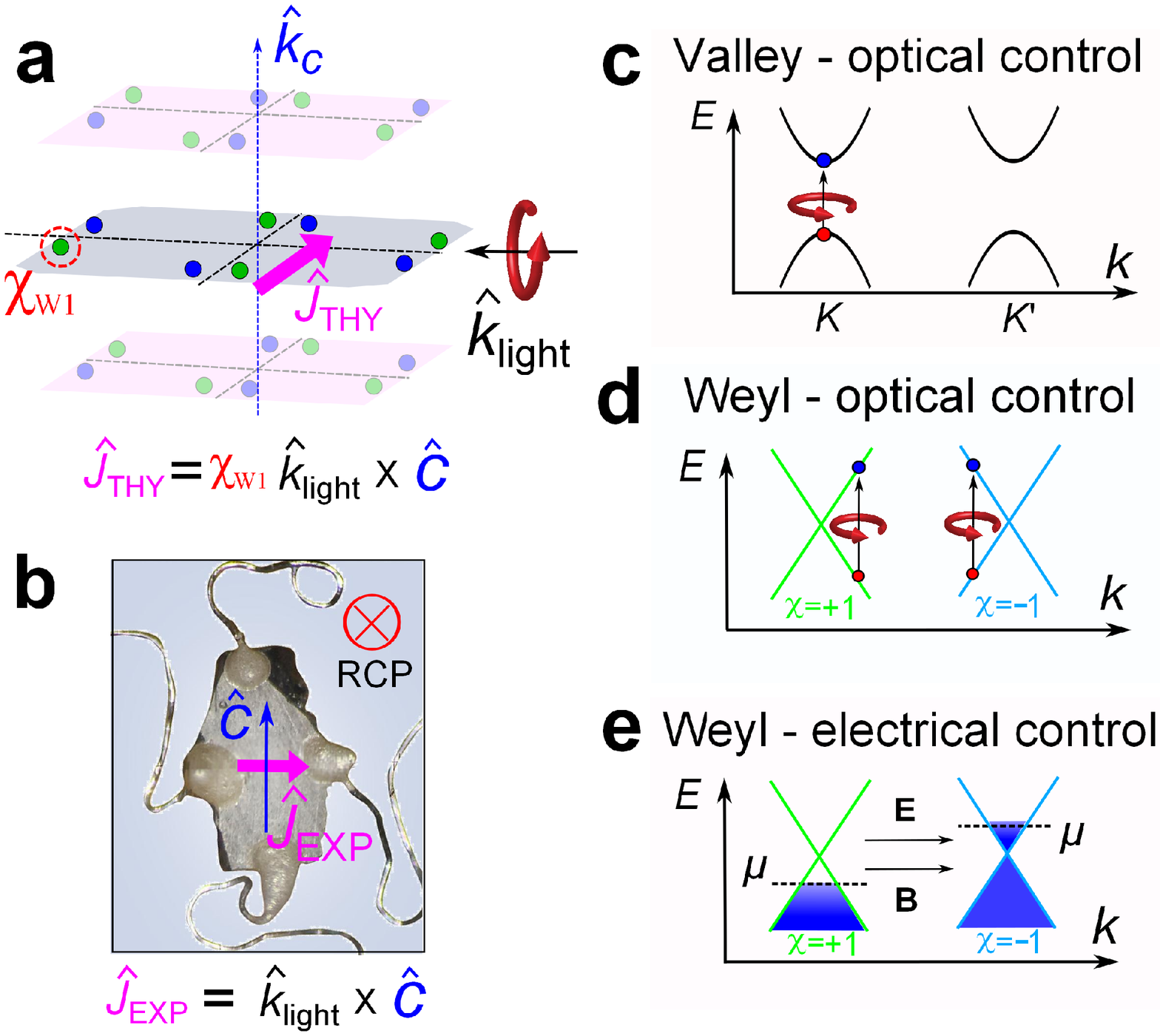}
\caption{ {\bf Detection and manipulation of chiral Weyl fermions by optical means.} \textbf{a,} Our calculations (see SI.II.3) show $\hat{J}_{\textrm{THY}}=\chi_{\textrm{W}_1}\hat{k}_{\textrm{light}}\times\hat{c}$. \textbf{b,} By measuring the direction of the current $\hat{J}$ and knowing the polarization and propagation direction of the light, we obtain $\hat{J}_{\textrm{EXP}}=\hat{k}_{\textrm{light}}\times\hat{c}$ from data. By comparing theory and data, we obtain $\chi_{\textrm{W}_1}=+1$. The $\chi_{\textrm{W}_1}=+1$ determined by the photocurrent agrees with that predicted by first-principles (Fig.~\ref{Fig1}\textbf{j}), further confirming our experimental detection of WF chirality. \textbf{c-e,} Comparison between the chirality degree of freedom of the WFs and valley degree of freedom in gapped Dirac system. \textbf{c,} In a gapped Dirac system, an optical excitation with a particular handedness can only populate one valley. \textbf{d,} In the Weyl system, an optical excitation with a particular handedness populates only one side of a single Weyl node. \textbf{e,} By applying parallel electric and magnetic fields, electrons can be pumped from one Weyl cone to the other of opposite chirality due to the chiral anomaly.}
\label{Fig4}
\end{figure*}

\end{document}